\documentclass[reprint,aps,graphicx,pre,showpacs]{revtex4-1}

\usepackage{amsmath,amssymb,amsfonts,graphicx,bm}
\usepackage{pgfplots}
\usepackage{epstopdf}
 \usepackage{paralist}
\usepackage[caption=false]{subfig}

\usepackage{amsmath,amsfonts,bm,bbm,MnSymbol}

\usepackage{float}
\usepackage{epsfig}
\usepackage{esint}
\usepackage{color}

\usepackage{multirow}

  \usepackage{paralist}
  \usepackage{epstopdf}
  \usepackage{graphics} 
 \usepackage[colorlinks=true]{hyperref}
 \hypersetup{urlcolor=blue, citecolor=red}
 \usepackage[latin1]{inputenc}
\usepackage[caption=false]{subfig}

 \usepackage{tikz-cd}

\usepackage{float}
\usepackage{epsfig}
\usepackage{esint}

\usepackage{bm,braket}

\newcommand{\bbT}{{\mathcal T}}

\newcommand{\q}{\widetilde{q}}

\newcommand{\R}{{\mathbb R}}

\newcommand{\e}{{\rm e}}

\newcommand{\calJ}{{\mathcal J}}
\newcommand{\calS}{{\mathcal S}}

\newcommand{\calP}{{\mathcal P}}

\newcommand{\calR}{{\mathcal R}}

\newcommand{\calM}{{\mathcal M}}

\newcommand{\x}{\mathbf{x}}
\newcommand{\y}{\mathbf{y}}

\newcommand{\X}{\mathbf{X}}
\renewcommand{\P}{\mathbb{P}}

\newcommand{\n}{\mathbf n}

\begin{document}
\title{Entropy production for diffusion processes across a semipermeable interface}
\author{Paul C. Bressloff}
\address{Department of Mathematics, Imperial College London, London SW7 2AZ, UK.}

\date{\today}

\begin{abstract}
The emerging field of stochastic thermodynamics extends classical ideas of entropy, heat and work to non-equilibrium systems. One notable finding is that the second law of thermodynamics typically only holds after taking appropriate averages with respect to an ensemble of stochastic trajectories. The resulting average rate of entropy production then quantifies the degree of departure from thermodynamic equilibrium. In this paper we investigate how the presence of a semipermeable interface increases the average entropy production of a single diffusing particle. Starting from the Gibbs-Shannon entropy for the particle probability density, we show that a semipermeable interface or membrane $\calS$ increases the average rate of entropy production by an amount that is equal to the product of the flux through the interface and the logarithm of the ratio of the probability density on either side of the interface, integrated along $\calS$. The entropy production rate thus vanishes at thermodynamic equilibrium, but can be nonzero during the relaxation to equilibrium, or if there exists a nonzero stationary equilibrium state (NESS). We illustrate the latter using the example of diffusion with stochastic resetting on a circle, and show that the average rate of interfacial entropy production is a nonmonotonic function of the resetting rate and the permeability. Finally, we give a probabilistic interpretation of the interfacial entropy production rate using so-called snapping out Brownian motion. This also allows us to construct a stochastic version of entropy production.

\end{abstract}
\maketitle


\section{Introduction}

In recent years there has been a rapid growth of interest in stochastic thermodynamics, which uses tools from the theory of stochastic processes to extend classical ideas of entropy, heat and work to non-equilibrium systems \cite{Sekimoto10,Seifert12,Cocconi20,Peliti21}. Examples include overdamped colloidal particles, biopolymers, enzymes, and molecular motors. One characteristic feature of such systems is that the second law of thermodynamics typically only holds after taking appropriate averages with respect to an ensemble of stochastic trajectories or over long time intervals. The resulting average rate of entropy production then
quantifies the degree of departure from thermodynamic equilibrium. In addition, probabilistic methods such as It\^o stochastic calculus, path integrals and Radon-Nikodym derivatives have been used to derive a variety of important fluctuation relations from the stochastic entropy evaluated along individual trajectories  \cite{Seifert05,Chetrite08,Sevick08,Jarzynski11}. These fluctuation relations have subsequently been generalized using martingale theory \cite{Chetrite11,Pigliotti17,Neri17,Neri19,Gal21,Roldan24}.

In this paper we consider the following problem: how does the presence of a semipermeable interface contribute to the average entropy production rate of a single diffusing particle? Diffusion through semipermeable membranes has a wide range of applications, including molecular transport through biological membranes  \cite{Philips12,Nik21,Bressloff22}, diffusion magnetic resonance imaging (dMRI) \cite{Tanner78,Coy94,Grebenkov10},  drug delivery \cite{Pontrelli07,Todo13}, reverse osmosis \cite{Li10}, and animal dispersal in heterogeneous landscapes \cite{Beyer16,Assis19,Kenkre21}. At the macroscopic level, the classical boundary condition for a semipermeable membrane takes the particle flux across the membrane to be continuous and to be proportional to the difference in concentrations on either side of the barrier. The constant of proportionality $\kappa_0$ is known as the permeability. The semipermeable boundary conditions are a particular version of the thermodynamically-derived Kedem-Katchalsky (KK) equations \cite{Kedem58,Kedem62,Kargol96,Aho16}. At the single-particle level, the resulting diffusion equation can be reinterpreted as the Fokker-Planck (FP) equation for the particle probability density, which is supplemented by the interfacial boundary conditions.  In particular, suppose that a semipermeable interface $\calS$ partitions $\R^d$ into two complementary domains $\Omega_{\pm}$ with $\R^d=\Omega_+\cup \Omega_-\cup \calS$. If $\calJ(\y,t)$ denotes the continuous flux across a point $\y\in \calS$ from $\Omega_-$ to $\Omega_+$, then $\calJ(\y, t)=\kappa_0\ [p_(\y^-,t)-p(\y^+,t)]/2$, where $p(\y^{\pm},t)$ are the solutions on $\calS_{\pm}=\overline{\Omega_{\pm}}\cap \calS$, where $\overline{\Omega}$ denotes closure of a set $\Omega$.

Starting from the Gibbs-Shannon entropy for the particle probability density $p(\x,t)$, $\x\in \R^d$, we show in Sect. II that a semipermeable interface $\calS $ increases the average rate of entropy production at a given time $t$ by an amount ${\mathcal I}_{\rm int}(t)=\int_{\calS}\calJ(\y,t)\ln[p(\y^-,t)/p(\y^+,t)]d\y$. In other words, ${\mathcal I}_{\rm int}(t)$ is equal to the product of the flux through the interface and the logarithm of the ratio of the probability density on either side of the interface, integrated along $\calS$. It immediately follows that the entropy production rate vanishes at thermodynamic equilibrium, but can be nonzero during the relaxation to equilibrium. We illustrate the theory by calculating the average rate of entropy production for one-dimensional (1D) diffusion. In the 1D case, $\calS$ reduces to a single point $x=0$, say, so that $\Omega_-=(-\infty,0^-]$ and $\Omega_+=[0^+,\infty)$.

In Sect. III we turn to a well-studied mechanism for maintaining a diffusing particle out of thermodynamic equilibrium, namely, stochastic resetting. Resetting was originally introduced within the context of a Brownian particle whose position $\X(t)$ instantaneously resets to a fixed position $\x_0$ at a sequence of times generated from a Poisson process of constant rate $r$  \cite{Evans11a,Evans11b,Evans14}. There have subsequently been a wide range of generalizations at the single particle level, see the review \cite{Evans20} and references therein. One of the signature features of diffusion processes with resetting is that there typically exists a nonequilibrium stationary state (NESS) that supports nonzero time-independent  fluxes. A number of recent studies have considered the stochastic thermodynamics of diffusive systems with resetting \cite{Fuchs16,Pal17,Busiello20,Gupta20,Pal21,Boyer21,Boyer22}. One issue that emerges from these studies is that sharp resetting to a single point is a unidirectional process that has
no time-reversed equivalent. This means that the average rate of entropy production calculated using the Gibbs-Shannon entropy cannot be related to the degree of time-reversal symmetry breaking. This connection can be established by considering resetting to a random position \cite{Mori23} or Brownian motion in an intermittent harmonic confining potential \cite{Alston22}.  An additional subtle feature arises when considering the effects of resetting in the presence of a semipermeable interface \cite{Bressloff22b,Grebenkov22,Bressloff23a}. In particular, it is natural to assume that the interface screens out resetting, in the sense that a resetting event cannot cross the interface. This means that a particle on one side of the interface $\partial \calM$ cannot reset to a point on the other side. Hence, it is not possible to have a non-zero stationary flux across the interface, since there is no countervailing reset current in the opposite direction. We bypass the screening effect in Sect. III by considering the example of single-particle diffusion on a ring with both stochastic resetting and a semipermeable interface. We derive an explicit expression for the resulting NESS and use this to calculate the various contributions to the average rate of entropy production in the stationary state, including those associated with resetting as well as those arising from the semipermeable interface.  

Finally, in Sect. IV we present a probabilistic interpretation of ${\mathcal I}_{\rm int}$  based on so-called snapping out BM \cite{Lejay16,Lejay18,Bressloff22a,Bressloff23}. The latter generates individual stochastic trajectories of the dynamics by sewing together successive rounds of partially reflected BMs that are restricted to either the left or right of the barrier.  Each round is killed (absorbed) at the barrier when its boundary local time exceeds an exponential random variable parameterized by the permeability $\kappa_0$. (The local time is a Brownian functional that specifies the contact time between a particle and a given boundary  \cite{Levy40,Ito63,Dynkin65,McKean75,Grebenkov06}.) A new round is then immediately started in either direction with equal probability. It is the random switching after each killing event that is the source of the entropy production. We also use snapping out BM to construct a stochastic version of entropy production. Averaging the latter with respect to the distribution of sample paths recovers the results based on the Gibbs-Shannon entropy.

\vfill

\section{Single-particle diffusion across a semipermeable interface}

 \begin{figure}[b!]
  \centering
  \includegraphics[width=8cm]{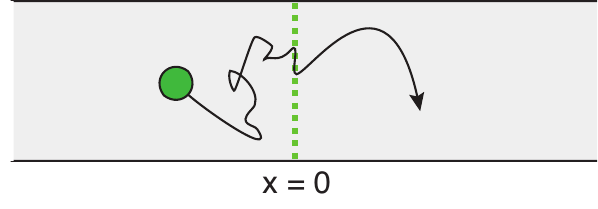}
  \caption{Brownian motion in $\R$ with a semipermeable interface at $x=0$. (The two-dimensional representation is for illustrative purposes.)}
  \label{fig1}
\end{figure}

\subsection{Diffusion in $\R$} Consider an overdamped Brownian particle diffusing in a 1D domain with a semipermeable barrier or interface at $x=0$. Suppose that the particle is also subject to a force $F(x,t)$. Let $p(x,t)$ denote the probability density of the particle at position $x$ at time $t$. The corresponding FP equation takes the form
\begin{subequations}
\label{rho1D}
\begin{equation}
\frac{\partial p(x,t)}{\partial t}=-\frac{\partial J(x,t)}{\partial x},\quad x\neq 0 ,\quad t>0,
\end{equation}
with the probability flux
\begin{equation}
J(x,t)=-D\frac{\partial p(x,t)}{\partial x}+\frac{1}{\gamma}F(x,t) p(x,t),
\end{equation}
and the following pair of boundary conditions at the interface:
\begin{equation}
J(0^{\pm},t)=\calJ(t):=\frac{\kappa_0}{2}[p(0^-,t)-p(0^+,t)],
\end{equation}
\end{subequations}
where $\kappa_0$ is a constant permeability. (The arbitrary factor of $1/2$ on the right-hand side of Eq. (\ref{rho1D}c) is motivated by the corresponding probabilistic interpretation of snapping out BM, see Sect. IV.) In addition, $D$ is the diffusivity, $\gamma$ is the friction coefficient, and the two quantities are related according to the Einstein relation $D\gamma=k_BT$. (In the following we set the Boltzmann constant $k_B=1$.)  
For simplicity, we take the diffusive medium to be spatially homogeneous. However, the domains $(-\infty,0^-]$ and $[0^+,\infty)$ could have different diffusivities, for example. That is, $D=D_-$ for $x<0$ and $D=D_+$ for $x>0$ with $D_-\neq D_+$; the drag coefficients would also differ due to the Einstein relation. Finally, we assume that the force $F(x,t)$ is continuous at the interface. 

In order to evaluate various thermodynamic quantities such as the average heat, work and entropy, we have  to integrate with respect to $x \in \R$. Since the probability density has a discontinuity at $x=0$, we partition each integral into the two domains $(-\infty,0^-]$ and $[0^+,\infty)$, and introduce the notation
\begin{equation}
\label{fint}
\fint dx=\int_{-\infty}^0dx+\int_0^{\infty}dx.
\end{equation}
Continuity of the flux at $x=0$ means that in most cases there is no contribution to the integral from the discontinuity. One notable exception occurs when we evaluate the average rate of entropy production.

In the 1D case we can always write the force as a gradient of a potential, $F(x,t)=-\partial_xV(x,t)$. This means that the average internal energy is
\begin{equation}
{\mathcal E}(t)=\fint dx\, p(x,t)V(x,t)  ,
\end{equation}
and
\begin{align}
\frac{d{\mathcal E}}{dt}&=\fint dx\,\left [ \frac{\partial p(x,t)}{\partial t}V(x,t)+ \frac{\partial V(x,t)}{\partial t}p(x,t) \right ].
\end{align}
Using Eqs. (\ref{rho1D}a,b) and integration by parts, we have
\begin{align}
&\int_{-\infty}^0dx\, \frac{\partial p(x,t)}{\partial t}V(x,t)\nonumber \\
&=\int_{-\infty}^0dx\,  \frac{\partial V(x,t)}{\partial x}J(x,t) -J(0^-,t)V(0,t)
\end{align}
and
\begin{align}
&\int_{0}^{\infty} dx\,  \frac{\partial p(x,t)}{\partial t}V(x,t)\nonumber\\
&= \int_{0}^{\infty} dx\,  \frac{\partial V(x,t)}{\partial t}J(x,t) +J(0^+,t)V(0,t).
\end{align}
Imposing flux continuity at the interface leads to a nonequilibrium version of the first law of thermodynamics:
\begin{equation}
\frac{d{\mathcal E}}{dt}=\frac{d{\mathcal W}}{dt}-\frac{dQ}{dt},
\end{equation}
where ${\mathcal W}(t)$ is the average work done on the particle and $Q(t)$ is the average heat dissipated into the environment with
\begin{align}
\frac{dQ}{dt}&=-\fint dx\,  \frac{\partial V(x,t)}{\partial x}J(x,t)  ,
\label{Q}
\end{align}
and
\begin{align}
\frac{d{\mathcal W}}{dt}&= \fint dx\, \frac{\partial V(x,t)}{\partial t}p(x,t) .
\end{align}
If the potential is time-independent then we have the further simplification that there exists a unique equilibrium stationary state given by the Boltzmann-Gibbs distribution,
\begin{equation}
\lim_{t\rightarrow \infty}p(x,t) =\frac{1}{Z} \e^{-V(x)/T},\quad \lim_{t\rightarrow \infty}J(x,t)=0.
\end{equation}
Since there are no fluxes at equilibrium, the semipermeable membrane becomes invisible, that is, it has no effect on the stationary state.

\begin{figure*}[t!]
  \centering
  \includegraphics[width=16cm]{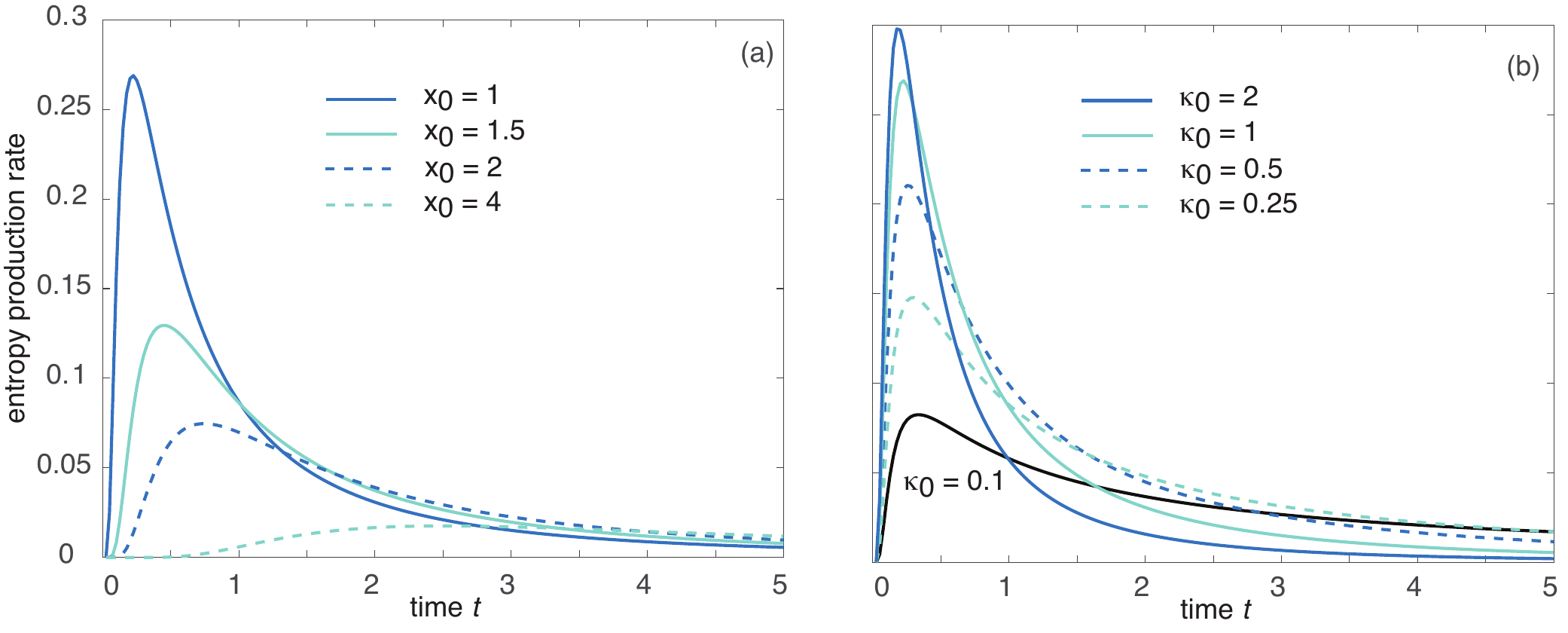}
  \caption{Single-particle diffusion across a closed semipermeable interface in $\R$. Plots of the average interfacial entropy production rate as a function time for (a) various resetting positions $x_0$ and fixed permeability $\kappa_0=1$ and (b) various $\kappa_0$ for fixed $x_0=1$.}
  \label{fig2}
\end{figure*}

\subsection{Average entropy production}

The average system entropy at time $t$ is defined by
\begin{align}
\label{GS1}
\calS^{\rm sys}(t):&= -\fint  dx\, p(x,t) \ln p(x,t),
\end{align}
which takes the form of a Gibbs-Shannon entropy.
In order to calculate the average rate of entropy production, we differentiate both sides of Eq. (\ref{GS1}) with respect to time $t$:
\begin{align}
\calR^{\rm sys}(t)&:=\frac{d\calS^{\rm sys}(t)}{dt}= -\fint  dx\, \frac{\partial p(x,t)}{\partial t}[1+ \ln p(x,t)].
\label{Rsys}
\end{align}
Using the FP equation and performing an integration by parts, we have
\begin{align}
&\calR^{\rm sys}(t)=\fint dx\, \frac{\partial J(x,t)}{\partial x}[1+ \ln p(x,t)]\nonumber \\
\label{Rsys2}
&=-\fint  dx\, \frac{J(x,t)}{p(x,t)}\frac{\partial p(x,t)}{\partial x}\\
& \quad +J(0^-,t))[1+\ln p(0^-,t)]-J(0^+,t))[1+\ln p(0^+,t)].\nonumber
\end{align}
Using the definitions of the probability fluxes, the integrand can be rewritten as
\begin{align}
&-\frac{1}{p(x,t)}\frac{\partial p(x,t)}{\partial x} J(x,t)
=\frac{J(x,t)^2}{D p(x,t)}- \frac{ F(x,t) J(x,t)}{T} .
\end{align}
In addition, imposing the permeability conditions shows that
\begin{align}
&J(0^-,t))[1+\ln p(0^-,t)]-J(0^+,t))[1+\ln p(0^+,t)] \nonumber \\
&\qquad 
=\calJ(t)\ln [p(0^-,t)/p(0^+,t)].
\end{align}
We thus obtain a generalization of the classical entropy production rate given by
\begin{align}
\calR^{\rm tot}(t) &:=\calR^{\rm sys}(t)+\calR^{\rm env}(t)\nonumber \\
&=\fint dx\, \frac{J(x,t)^2}{D p(x,t)} +{\mathcal I}_{\rm int}(t),
\label{Rtot}
\end{align}
where 
\begin{equation}
\label{Iint}
 {\mathcal I}_{\rm int}(t)=\frac{\kappa_0}{2}[p(0^-,t)-p(0^+,t)]\ln \left [\frac{p(0^-,t)}{p(0^+,t)}\right ]
 \end{equation}
 is the contribution to the entropy production from the semipermeable membrane, and
\begin{align}
&\calR^{\rm env}(t):=\frac{1}{T}\frac{dQ}{dt}=\fint dx\, \frac{F(x,t)J(x,t)}{T}
\end{align}
is the average environmental entropy production rate due to heat dissipation, see Eq. (\ref{Q}). Since $[p(0^-,t)-p(0^+,t)]$ and $\ln[p(0^-,t)/\ln p(0^+,t)]$ have the same sign, it follows that $ {\mathcal I}_{\rm int}(t)\geq 0$, and hence the average total entropy production rate satisfies the second law of thermodynamics in the sense that
\begin{equation}
\calR^{\rm sys}(t)+\calR^{\rm env}(t)\geq 0,\quad t\geq 0.
\end{equation}
The existence of the contribution ${\mathcal I}_{\rm int}$ is one of the main results of our paper. We will give a physical interpretation of this result in Sect. IV.

\subsection{Interfacial entropy production for pure diffusion}

In the particular case of a time-independent force $F(x)=-V'(x)$, the rate of entropy production vanishes in the limit $t\rightarrow \infty$ since there are no fluxes at equilibrium. However, $\calR^{\rm tot}(t)>0$ at finite times $t$. This result holds even for pure diffusion, where
 an explicit solution of Eqs. (\ref{rho1D}) can be obtained. The simplest way to proceed is to Laplace transform Eqs. (\ref{rho1D}), under the initial condition $p(x,0)=\delta(x-x_0)$. For the sake of illustration we take $x_0>0$. It follows that 
\begin{equation}
\widetilde{p}(x,s):=\int_0^{\infty} \e^{-st}p(x,t)dt =G(x,x_0;s),
\end{equation}
where $G(x,_0;s)$ is the Green's function of the modified Helmholtz equation
\begin{subequations}
\label{G}
\begin{equation}
\frac{d^2G}{dx^2}-sG(x,x_0;s)=-\delta(x-x_0),
\end{equation}
supplemented by the interfacial conditions
\begin{align}
&-D\frac{dG(x,x_0;s)}{dx}|_{x=0^-}=-D\frac{dG(x,x_0;s)}{dx}|_{x=0^+}\nonumber \\
&=\frac{\kappa_0}{2}[G(0^-,x_0;s)-G(0^+,x_0;s)],
\end{align}
and with $\lim_{x\rightarrow\pm \infty}G(x,x_0;s)=0$. 
\end{subequations}
Eq. (\ref{G}a) has the general solution
\begin{equation}
G(x,x_0;s)=\frac{\e^{\sqrt{s/D}|x-x_0|}}{2\sqrt{sD}}+A(s)\e^{\sqrt{s/D}x}+B(s)\e^{-\sqrt{s/D}x},
\end{equation}
with the pair of coefficients $A(s)$ and $B(s)$ determined by the supplementary conditions.
Finally, inverting the resulting solution in Laplace space gives
\begin{subequations}
\begin{align}
&p(x,t) \\
&=\frac{1}{2\sqrt{\pi D t}}\left [\exp\left (-\frac{(x-x_0)^2}{4Dt}\right )+\exp\left (-\frac{(x+x_0)^2}{4Dt}\right )\right ]\nonumber \\
& -\frac{\kappa_0}{2D}\exp\left (\frac{\kappa_0}{D}(x+x_0 +\kappa_0t)\right )\mbox{erfc}\left (\frac{(x+x_0+2\kappa_0t}{2\sqrt{Dt}}\right ),\nonumber
\end{align}
for $x>0$ and
\begin{align}
&p(x,t) \\
&=\frac{\kappa_0}{2D}\exp\left (\frac{\kappa_0}{D}(x_0-x+\kappa_0t)\right )\mbox{erfc}\left (\frac{x_0-x+2\kappa_0t}{2\sqrt{Dt}}\right )\nonumber
\end{align}
\end{subequations}
for $x<0$.
The complementary error function is
\begin{equation}
\mbox{erfc}(x):=\frac{2}{\sqrt{\pi}}\int_x^{\infty} \e^{-y^2}dy
\end{equation}

In the limit $\kappa_0\rightarrow 0$, we see that $p(x,t)\rightarrow 0$ for $x<0$ and $p(x,t)\rightarrow p_+(x,t)$ for $x>0$, with
\begin{align}
&p_+(x,t)\\
&= \frac{1}{2\sqrt{\pi D t}}\left [\exp\left (-\frac{(x-x_0)^2}{4Dt}\right )+\exp\left (-\frac{(x+x_0)^2}{4Dt}\right )\right ].\nonumber
\end{align}
This is consistent with the fact that the interface becomes completely impermeable in the limit $\kappa_0\rightarrow 0$ and the particle started to the right of the interface. We thus recover the solution of the diffusion equation on the half-line with a totally reflecting boundary at $x=0$. In order to determine what happens in the limit $\kappa_0\rightarrow \infty$, we use
 the asymptotic expansion
\begin{equation}
\mbox{erfc}(x)\sim \frac{\e^{-x^2}}{\sqrt{\pi}x}\left [1-\frac{1}{2x^2}+\ldots \right ].
\end{equation}
We find that the interface becomes completely transparent and $p(x,t)$ is given by the classical solution of free diffusion in $\R$. In Fig. \ref{fig2} we show sample plots of the interfacial entropy production rate ${\mathcal I}_{\rm int}(t)$, see Eq. (\ref{Iint}), as a function of time $t$ for various initial positions $x_0$ and permeabilities $\kappa_0$. As expected, the effects of the interface at $x=0$ are greater when $x_0$ is closer to the origin. In addition, increasing the permeability $\kappa_0$ increases the initial rise of ${\mathcal I}_{\rm int}(t)$ but the effects of the interface decay more quickly.

\begin{figure}[b!]
  \centering
  \includegraphics[width=6cm]{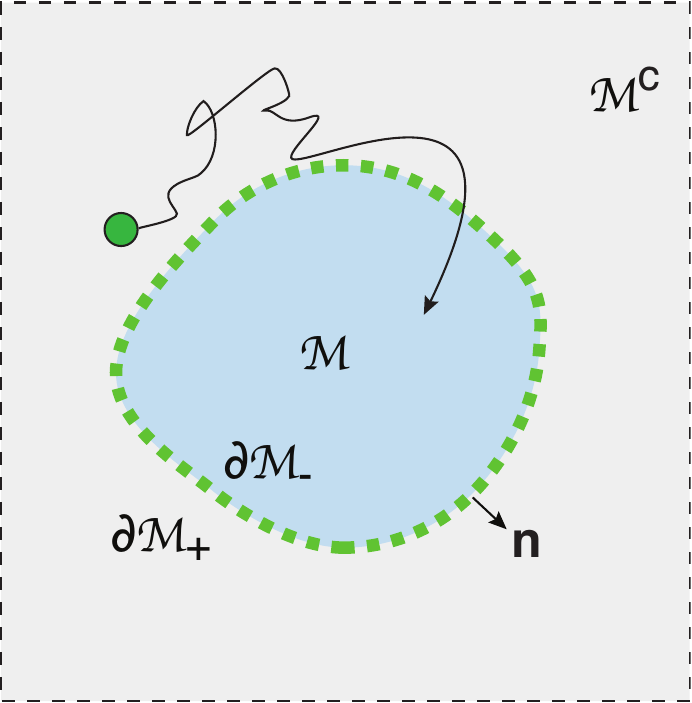}
  \caption{Single-particle diffusion across a closed semipermeable membrane in $\R^d$.}
  \label{fig3}
\end{figure}

\subsection{Diffusion across a closed semipermeable membrane in $\R^d$}

The expression (\ref{Iint}) for the average rate of entropy production through a semipermeable interface generalizes to higher spatial dimensions. Suppose that $\calM$ denotes a closed bounded domain $\calM\subset \R^d$ with a smooth concave boundary $\partial \calM$ separating the two open domains $\calM$ and its complement $ \calM^c=\R^d\backslash \calM$, see Fig. \ref{fig3}.  The boundary acts as a semipermeable interface with $\partial \calM_+ $ ($\partial \calM_-$) denoting the side approached from outside (inside) $\calM$. Let $p(\x,t)$ denote the probability density function of an overdamped Brownian particle subject to a force field ${\bf F}(\x)$. The multi-dimensional analog of the FP Eq. is
\begin{subequations}
\label{dclass}
\begin{align}
\frac{\partial p(\x,t)}{\partial t}&=-{\bm \nabla}\cdot {\bf J}(\x,t),\ \x \in\calM\cup \calM^c,\\
{\bf J}(\x,t)&=-D{\bm \nabla}p(\x,t)+\frac{1}{\gamma}{\bf F}(\x) p(\x,t),\\
 {\bf J}(\y^{\pm},t)\cdot {\bf n}&=\calJ(\y,t), \quad \y  \in \partial \calM,\\
\calJ(\y,t)&=\frac{\kappa_0}{2}[p(\y^-,t)-p(\y^+,t)],\quad \y \in \partial \calM,
\end{align}
\end{subequations}
where  $\n$ is the unit normal directed out of $\calM$.

The average system entropy at time $t$ is defined by
\begin{align}
\label{GSd}
\calS^{\rm sys}(t):&= -\int_{\calM} d\x\, p(\x,t) \ln p(\x,t)\nonumber \\
&\quad  -\int_{\calM^c} d\x\, p(\x,t) \ln p(\x,t).
\end{align}
Differentiating both sides of Eq. (\ref{GSd}) with respect to time $t$ gives
\begin{align}
\calR^{\rm sys}(t)&:=\frac{d\calS^{\rm sys}(t)}{dt}= -\int_{\calM} d\x\,\, \frac{\partial p(\x,t)}{\partial t}[1+ \ln p(\x,t)]\nonumber \\
&\qquad -\int_{\calM^c} d\x \, \frac{\partial p(\x,t)}{\partial t} [1+\ln p(\x,t)].
\label{dRsys}
\end{align}
Using the FP Eq. (\ref{dclass}) and performing an integration by parts according to the divergence theorem, we have
\begin{align}
&\calR^{\rm sys}(t)
= \int_{\calM} d\x\,     [1+ \ln p(\x,t)]{\bm \nabla}\cdot {\bf J}(\x,t)  \nonumber \\
&\qquad \qquad +\int_{\calM^c} d\x \,       [1+\ln p(\x,t)]{\bm \nabla}\cdot {\bf J}(\x,t)  \nonumber \\
&=-\int_{\calM} d\x\, {\bm \nabla}p(\x,t) \cdot \frac{{\bf J}(\x,t) }{p(\x,t)}     -\int_{\calM^c} d\x \,    {\bm \nabla}p(\x,t) \cdot \frac{{\bf J}(\x,t) }{p(\x,t)}  \nonumber\\
& \qquad+ \int_{\partial \calM_-} d\y\,  [1+ \ln p(\y,t)] {\bf J}(\y,t) \cdot \n  \nonumber \\
&\qquad -\int_{\partial \calM_+} d\y\, [1+ \ln p(\y,t)] {\bf J}(\y,t)   \cdot \n .
\end{align}
We now decompose the various contributions to $\calR^{\rm sys}$ along analogous lines to the 1D case. This leads to the higher dimensional version of Eq. (\ref{Rtot}):
\begin{align}
&\calR^{\rm tot}(t) :=\calR^{\rm sys}(t)+\calR^{\rm env}(t)\nonumber \\
&= \int_{ \calM} d\x\, \frac{{\bf J}(\x,t)^2}{D p(\x,t)}
+ \int_{\calM^c} d\x\, \frac{{\bf J}(\x,t)^2}{D p(\x,t)} +{\mathcal I}_{\rm int}(t),
\label{dRtot}
\end{align}
where 
\begin{equation}
\label{dIint}
 {\mathcal I}_{\rm int}(t)=\int_{\partial \calM} d\y\, \calJ(\y,t) \ln \left [\frac{p(\y^-,t)}{p(\y^+,t)}\right ]
 \end{equation}
 is the contribution to the entropy production from the semipermeable membrane, and
\begin{align}
&\calR^{\rm env}(t)=\frac{1}{T}\frac{dQ}{dt}\\
&= \int_{ \calM} d\x\, \frac{{\bf F}(\x,t)\cdot {\bf J}(\x,t)}{T}
+ \int_{ \calM^c} d\x\, \frac{{\bf F}(\x,t)\cdot {\bf J}(\x,t)}{T}. \nonumber 
\end{align}

 \subsection{Imperfect contacts and the chemical potential}
A classical generalization of the permeable boundary condition (\ref{rho1D}c), or its higher-dimensional generalization (\ref{dclass}c), is to include a directional asymmetry in the permeability, which can be interpreted as a step discontinuity in a chemical potential \cite{Kedem58,Kedem62,Kargol96,Farago20}:
\begin{equation}
\label{asym}
J(0^{\pm},t)=\calJ(t):=\frac{\kappa_0}{2}[p(0^-,t)-\sigma p(0^+,t)]
\end{equation}
for $0\leq \sigma \leq 1$. This tends to enhance the concentration to the right of the interface. (If $\sigma >1$ then we would have an interface with permeability $\kappa_0\sigma$ and bias $1/\sigma$ to the left.) The derivation of the average entropy production rate for a 1D diffusive process proceeds along similar lines as the symmetric case, leading to Eq. (\ref{Rtot}). However, the interfacial contribution  (\ref{Iint}) now takes the form 
\begin{align}
 {\mathcal I}_{\rm int}(t)&=\frac{\kappa_0}{2}[p(0^-,t)-\sigma p(0^+,t)]\ln \left [\frac{p(0^-,t)}{p(0^+,t)}\right ],
 \end{align}
 which is not necessarily positive. In order to obtain the correct second law of thermodynamics, we decompose ${\mathcal I}_{\rm int}(t)$ as
\begin{align}
 {\mathcal I}_{\rm int}(t)&={\mathcal I}_{\sigma}(t) +\frac{\kappa_0}{2}[p(0^-,t)-\sigma p(0^+,t)]\ln \sigma,
 \end{align}
 with
 \begin{align}
 {\mathcal I}_{\sigma}(t)&=\frac{\kappa_0}{2}[p(0^-,t)-\sigma p(0^+,t)]\ln \left [\frac{p(0^-,t)}{\sigma p(0^+,t)}\right ]\geq 0.
 \end{align}
 Suppose that there is a discontinuity in the chemical potential across the interface, with $\mu=\mu_-$ for $x<0$ and $\mu =\mu_+$ for $x>0$. We then make the identification $\sigma=\e^{(\mu_+-\mu_-)/T}$ with $\mu_+<\mu_-$ for $\sigma <1$, such that
 \begin{equation}
  {\mathcal I}_{\rm int}(t)={\mathcal I}_{\sigma}(t)+\calJ(t) \frac{\mu_+-\mu_-}{T},
  \end{equation}
  The second term on the right-hand side represents the rate of reduction in the free energy due to the probability flux $\calJ(t)$ from a region with a high chemical potential $\mu_-$ to a region with a low chemical potential $\mu_+$. This change in free energy contributes to the heat dissipated into the environment. Hence, redefining the environmental entropy according to 
\begin{align}
\calR^{\rm env}(t)
&=\fint dx\, \frac{F(x,t)J(x,t)}{T} +\calJ(t) \frac{\mu_--\mu_+}{T},
\end{align}
we obtain the modified second law of thermodynamics
\begin{align}
&\calR^{\rm tot}(t) 
=\fint dx\, \frac{J(x,t)^2}{D p(x,t)} +{\mathcal I}_{\sigma}(t)\geq 0.
\label{Rtot2}
\end{align}

\setcounter{equation}{0}

\section{Stochastic resetting on circle with a semipermeable interface}

We now turn to an example that supports an NESS in the large time limit by combining diffusion across a semipermeable interface with stochastic resetting. In previous work we considered scenarios similar to the one shown in Fig. 4(a) \cite{Bressloff22b,Bressloff23a}. We assumed that the semipermeable interface acts as a screen for resetting in the sense that a particle located in $\calM$ cannot reset to a point $\x_0\in \calM^c$ and vice versa. This means that although the NESS is characterized by non-zero stationary fluxes in both domains, the net flux across the interface is zero. The last result can be understood as follows. Suppose, for concreteness, that the stationary interfacial flux $\calJ^*$ is positive so that there is a net flow of probability from $\calM$ to $\calM^c$. The screening effect of the interface means that there is no reset flux in the opposite direction, which is impossible for a stationary state with $p^*(\x)>0$ unless $\calJ^*=0$. The vanishing of the interfacial flux can also be confirmed by explicitly calculating the NESS \cite{Bressloff22b,Bressloff23a}.

\begin{figure*}[t!]
  \centering
  \includegraphics[width=14cm]{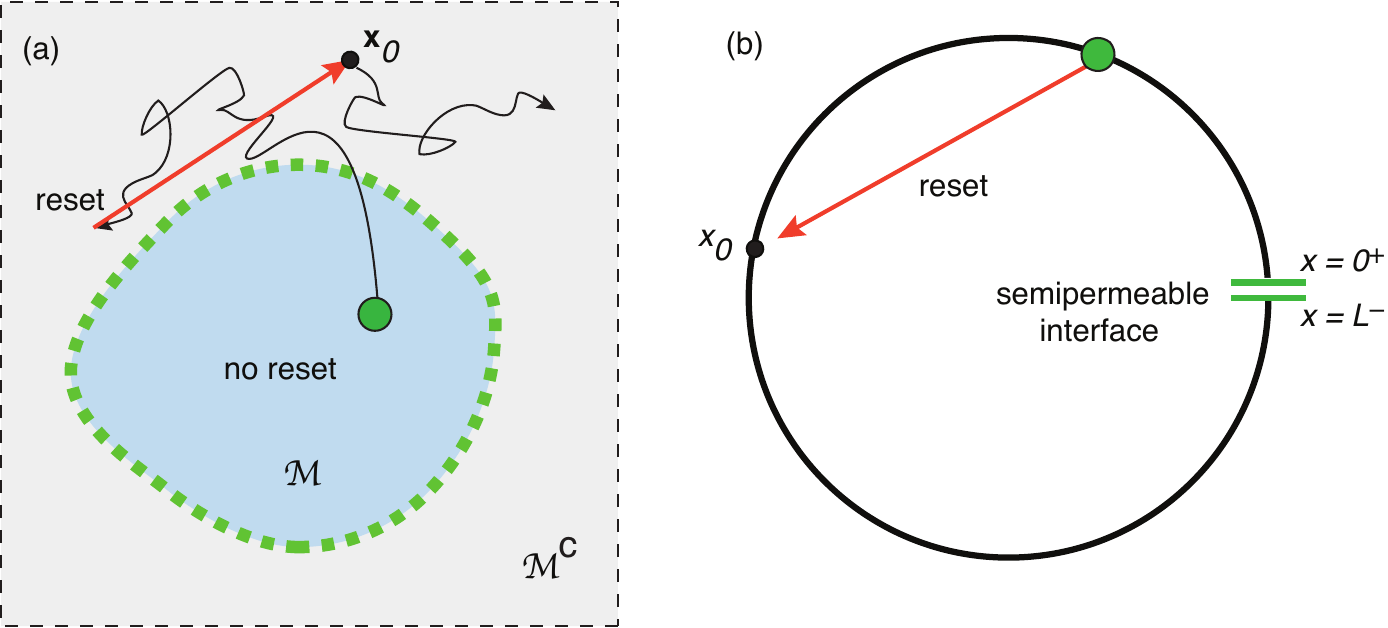}
  \caption{Screening effect of a semipermeable membrane for a diffusing particle with resetting. (a) Closed semipermeable membrane $\partial \calM$ in $\R^d$ with a resetting point $\x_0\in \calM^c$. Although the particle can diffuse across $\partial \calM$ in either direction, it cannot reset to $\x_0$ whenever it is within $\calM$. (b) Single particle diffusing on a circle with a semipermeable interface at $\{0,+,L^-\}$, and resetting to a point $x_0$. Let $X(t)$ denote the current particle position. If $X(t)\in [0^+,x_0)$ then it can only reset in the anticlockwise direction, otherwise it can only reset in the clockwise direction.  }
  \label{fig4}
\end{figure*}

Therefore, in contrast to our previous work, we consider a Brownian particle diffusing on a circle of circumference $L$. The circle is topologically equivalent to a finite interval $[0,L]$ with a semipermeable interface at $x=\{0^+,L^-\}$. The particle is taken to reset at a rate $r$ to a random position $y\in (0,L)$ generated from the density $\sigma_0(y)$. The presence of the semipermeable interface means that $\sigma_0(y)$ does not have to be a periodic function. In Fig. \ref{fig4}(b) we show the example of resetting to a single point $x_0$, which is obtained by taking $\sigma_0(y)=\delta(y-x_0)$. Since the particle cannot reset by crossing the semipermeable interface, it resets in the anticlockwise direction when $X(t)\in [0^+,x_0)$ and resets in the clockwise direction when $X(t)\in (x_0,L^-]$. For a random reset location,
the probability density $p(x,t)$, $x\in [0,L]$, evolves according to
 the equation
\begin{subequations}
\label{FPr}
\begin{align}
\frac{\partial p}{\partial t}&=D\frac{\partial^2 p(x,t)}{\partial x^2}-rp(x,t)+r\sigma_0(x),
\end{align}
together with the interfacial conditions
\begin{align}
-D\frac{\partial p(0^+,t)}{\partial x}&=-D\frac{\partial p(L^-,t)}{\partial x}=
\calJ(t),\\
\calJ(t)&= \kappa_0 [p(L^-,t)-p(0^+,t)].
\end{align}
\end{subequations}
(A factor of $1/2$ has been absorbed into $\kappa_0$.)

\subsection{Average entropy production}

We begin by calculating the average rate of entropy production, and show that there are contributions from the interface and the resetting protocol, We then calculate these contributions in the NESS. Substituting Eqs. (\ref{FPr}a) into the formula (\ref{Rsys}) for the average rate of entropy production gives
\begin{align}
\calR^{\rm sys}(t)&=\int_0^L dx\, \frac{\partial J(x,t)}{\partial x} [1+\ln p(x,t)] +{\mathcal I}_r(t),
\label{goo}
\end{align}
where $J(x,t)=-D\partial p(x,t)/\partial x$ and
\begin{align}
{\mathcal I}_r(t)&= r\int_0^L dx\, \
 [p(x,t)-\sigma_0(x)] \ln p(x,t) .
\end{align}
The integral on the right-hand side of Eq. (\ref{goo}) can be analyzed along identical lines to the case of no resetting. Since there are no forces, we find that
\begin{align}
\calR^{\rm sys}(t)
&=\int_0^{L}dx\,  \frac{J(x,t)^2}{D p(x,t)}+ {\mathcal I}_{\rm int}(t)+r {\mathcal I}_r (t),\label{ent0}\end{align}
with
\begin{equation}
\label{Iint2}
 {\mathcal I}_{\rm int}(t)= \kappa_0 [p(L^-,t)-p(0^+,t)]\ln \left [\frac{p(L^-,t)}{p(0^+,t)}\right ].
 \end{equation}
Hence, there are contributions from both the semipermeable interface and resetting.

In the special case $\sigma_{0}(x)=\delta(x-x_0)$ (resetting to a fixed location $x_0$), we have
\begin{equation}
{\mathcal I}_r (t)=r\int_0^Ldx\, p(x,t)\ln p(x,t)-r\ln p(x_0,t).
\end{equation}
Following along analogous lines to Ref. \cite{Seifert12} we define ${\mathcal R}^{\rm reset}(t)=- {\mathcal I}_r (t) $ and rewrite Eq. (\ref{ent0}) as
\begin{align}
&\calR^{\rm sys}(t)+{\mathcal R}^{\rm reset}(t)
=\int_0^{L} dx\, \frac{J(x,t)^2}{D p(x,t)} +{\mathcal I}_{\rm int}(t)\geq 0.
\end{align}
As highlighted in the introduction, resetting to a single point is a unidirectional process that has
no time-reversed equivalent. This means that the average rate of entropy production calculated using the Gibbs-Shannon entropy cannot be related to the degree of time-reversal symmetry breaking. 
On the other hand, as shown in Ref. \cite{Mori23}, such a connection can be made in the case of a regular resetting density $\sigma_{0}(x)>0$. The contribution ${\mathcal I}_r (t)$ is now decomposed as
\begin{align}
{\mathcal I}_r (t)&=K[p(\cdot,t)|\sigma_{0}]+K[\sigma_{0}|p(\cdot,t)]\nonumber \\
&\quad - \int_0^L dx \,[ \sigma_0(x)-p(x,t)] \ln \sigma_0(x),
\end{align}
where $K[p|q]$ is the Kullback-Leibler divergence of any two measures $p,q$ on $\R$:
\begin{equation}
K[p|q]=\int dx \, p(x)\ln[p(x)/q(x)].
\end{equation}
Using Jensen's inequality for convex functions it is straightforward to prove that $K[p|q]\geq 0$. Hence, we can rewrite the entropy production equation as
 \begin{align}
&\calR^{\rm sys}(t)+\calR^{\rm res}(t)\nonumber \\
&= \int_0^{L} \frac{J(x,t)^2}{D p(x,t)}+{\mathcal I}_{\rm int}(t) +r (K[p(\cdot,t)|\sigma_{0}]+K[\sigma_{0}|p(\cdot,t)])\nonumber \\
& \geq 0,
\label{kk}
\end{align}
where
\begin{align}
\calR^{\rm res}(t) &= r\int_{0}^{L}dx\, [\sigma_0(x)-p(x,t)]\ln \sigma_0(x).\nonumber
\end{align}

\subsection{Non-equilibrium stationary state (NESS)}

In contrast to the system without resetting, there exists a nonequilibrium stationary state (NESS) for which there are non-zero stationary fluxes $J^*(x)$ and $\calJ^*$, both of which contribute to the average rate of entropy production. 
Setting all time derivatives to zero in Eqs. (\ref{FPr}) gives
\begin{subequations}
\label{ness}
\begin{align}
 &D\frac{d^2p^*(x)}{d x^2} -rp^*(x)=-r \sigma_0(x), \\
&-D\frac{d p^*(0)}{d x}=-D\frac{d p^*(L)}{d x}=
\calJ^*=\kappa_0[p^*(L)-p^*(0)].
\end{align}
\end{subequations}
For the moment, suppose that $\sigma_0(x)=\delta(x-x_0)$. The general solution of Eq. (\ref{ness}a) then takes the form
\begin{subequations}
\label{solas}
\begin{align}
p^*=p_-^*(x,x_0)&=A_1(x_0)\e^{\eta x}+B_1(x_0)\e^{-\eta x},\quad 0<x<x_0\\
p^*=p_+^*(x,x_0)&=A_2(x_0)\e^{\eta x}+B_2(x_0)\e^{-\eta x},\quad x_0<x<L,
\end{align}
\end{subequations}
with $\eta=\sqrt{r/D}$. There are four unknown coefficients and four supplementary conditions. The first pair arises from the conditions at $x_0$:
\begin{subequations}
\begin{align}
&p_-^*(x_0,x_0)=p_+^*(x_0,x_0),\\  & \left . D\frac{dp_+^*(x,x_0)}{dx}\right  |_{x=x_0}-\left . D\frac{dp_-^*(x,x_0)}{dx}\right  |_{x=x_0}=-r/D,
\end{align}
\end{subequations}
that is
\begin{subequations}
\label{ABa}
\begin{align}
&A_2\e^{\eta x_0}+B_2\e^{-\eta x_0}=A_1\e^{\eta x_0}+B_1\e^{-\eta x_0},\\
&A_2\e^{\eta x_0}-B_2\e^{-\eta x_0}=A_1\e^{\eta x_0}-B_1\e^{-\eta x_0}-\frac{r}{\eta D}.
\end{align}
\end{subequations}
The second pair follow from the interfacial conditions (\ref{ness}b):
\begin{subequations}
\label{ABb}
\begin{align}
&A_2\e^{\eta L}-B_2\e^{-\eta L}=A_1-B_1 ,\\
&A_2\e^{\eta L}+B_2\e^{-\eta L}=\left (1-\frac{D\eta}{\kappa_0}\right )A_1 +\left (1+\frac{D\eta}{\kappa_0}\right )B_1  .
\end{align}
\end{subequations}
After some algebra we find that
\begin{subequations}
\label{coeff}
\begin{align}
A_1&=\frac{(a_-+b_-) \Gamma_-(x_0)\e^{\eta L}-(a_--b_-)\Gamma_+(x_0)\e^{-\eta L}}
{b_-a_++b_+a_-},\\
B_1&=\frac{(a_+-b_+) \Gamma_-(x_0)\e^{\eta L}-(a_++b_+)\Gamma_+(x_0)\e^{-\eta L}}
{b_-a_++b_+a_-},\end{align}
and
\begin{equation}
A_2(x_0)=A_1(x_0)-\Gamma_-(x_0),\quad B_2(x_0)=B_1(x_0)+\Gamma_+(x_0),
\end{equation}
\end{subequations}
where
\begin{align}
a_{\pm} =\e^{\pm \eta L}-1,\quad b_{\pm} =a_{\pm}\pm \frac{D\eta}{\kappa_0},\quad \Gamma_{\pm}(x_0)=\frac{r}{2\eta D} \e^{\pm \eta x_0}.
\end{align}

 \begin{figure*}[t!]
\centering
 \includegraphics[width=16cm]{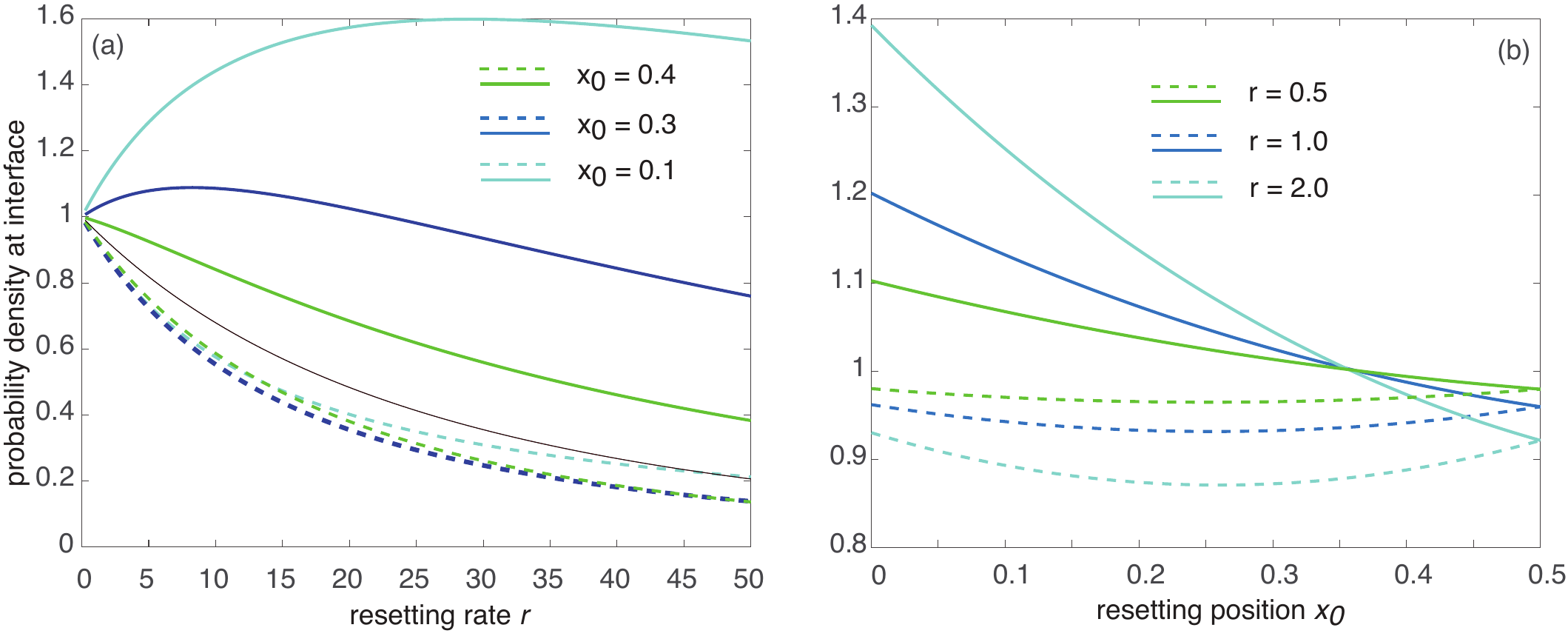}
  \caption{Single-particle diffusion on a ring with resetting to the same location $x_0$. The ring is mapped to the interval $ [0,L]$ with a semipermeable interface at $x=\{0^+,L^-\}$. Plots of the stationary densities $p^*(0^+)$ (solid curves) and $p^*(L^-)$ (dashed curves) on either side of the interface as a function of (a) the resetting rate $r$ for fixed $x_0$ and (b) the resetting position $x_0$ for fixed $r$. Other parameter values are $D=\kappa_0=L=1$. The thin solid line in (a) is the stationary density when $x_0=0.5$; by symmetry the density is continuous across the interface.}
  \label{fig5}
\end{figure*}

 \begin{figure*}[t!]
\centering
 \includegraphics[width=16cm]{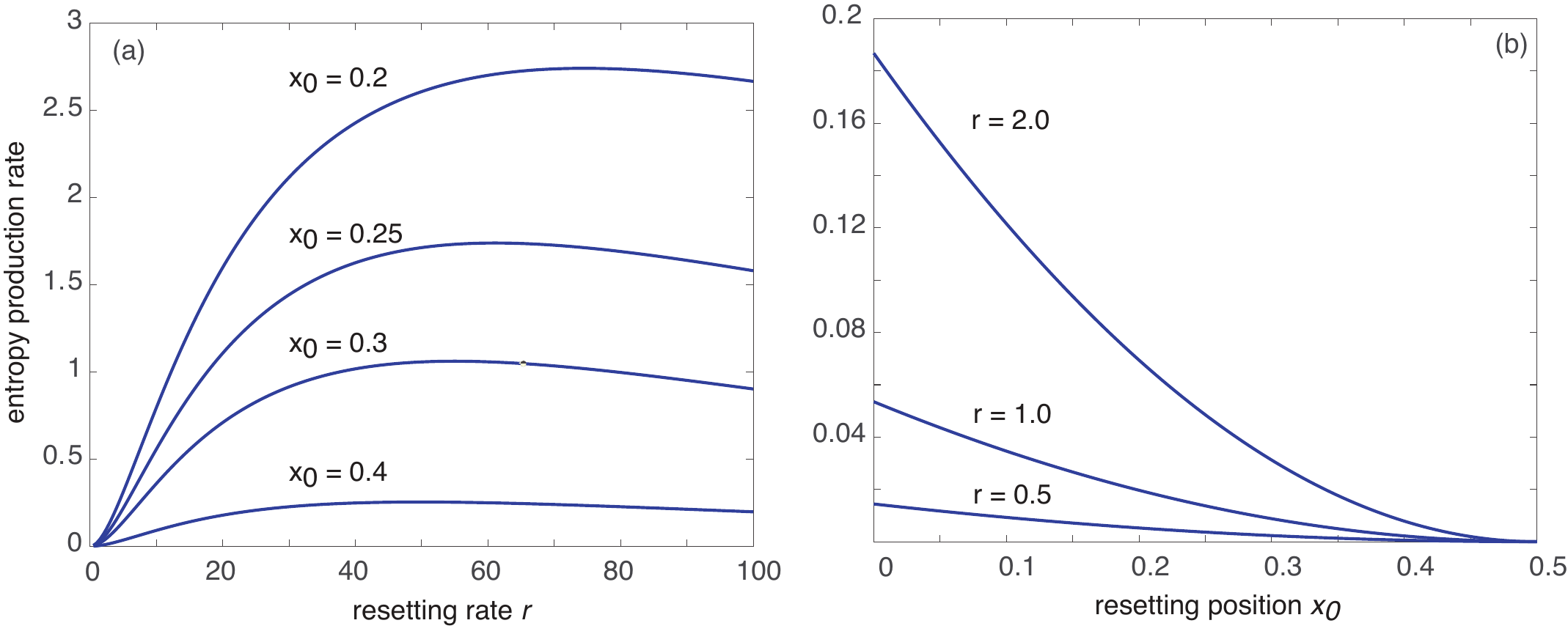}
  \caption{Single-particle diffusion on a ring with resetting to the same location $x_0$. Plots of the average rate of interfacial entropy production ${\mathcal I}_{\rm int}^*$ as a function of (a) the resetting rate $r$ and (b) the resetting location $x_0$. Other parameter values are the same as Fig. \ref{fig5}.}
  \label{fig6}
\end{figure*}

 \begin{figure}[t!]
\centering
 \includegraphics[width=8cm]{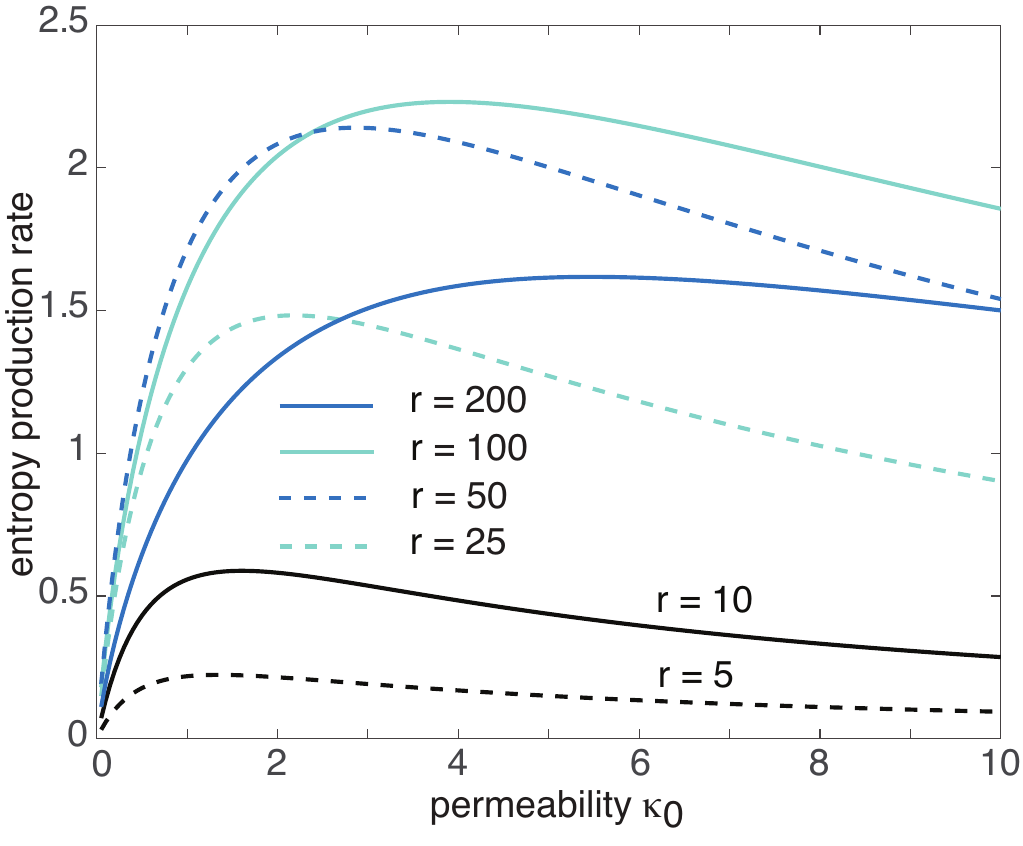}
  \caption{Single-particle diffusion on a ring with resetting to the same location $x_0=0.25$. Plots of the average rate of interfacial entropy production ${\mathcal I}_{\rm int}^*$ as a function of the permeability $\kappa_0$ and different values of $r$. Other parameter values are the same as Fig. \ref{fig5}.}
  \label{fig7}
\end{figure}

 In Figs. \ref{fig5}(a,b) we plot the solutions $p_-^*(0,x_0)$ and $p_+^*(L,x_0)$ on either side of the semipermeable interface as a function of $r$ and $x_0$, respectively. We fix the space and time units by setting $L=D=1$. A number of observations can be made. First, if $x_0=0.5 $ then $p_-^*(0,x_0)=p_+^*(L,x_0)$ for all resetting rates $r\geq 0$. This is a consequence of the symmetry of the configuration under the reflection $x\rightarrow 1-x$. This symmetry no longer holds when $x_0\neq 0.5$ and $r>0$. That is, $p_-^*(0,x_0)>p_+^*(L,x_0)$ for $x_0\in (0,0.5)$ and there is now an exchange symmetry $p_-^*(0,x_0)\leftrightarrow p_+^*(L,x_0)$ under reflection. Second, for fixed $x_0\in (0,L)$, the discontinuity across the interface is a nonmonotonic function of $r$, since the stationary state is at equilibrium when $r=0$ and the density at the interface vanishes in the limit $r\rightarrow \infty$. On the other hand, the discontinuity is a monotonically decreasing function of $x_0\in (0,0.5)$. These results are reflected in plots of the average entropy production rate at the interface, 
 \begin{equation}
 {\mathcal I}_{\rm int}^*=\calJ^*(x_0)\ln [p_-^*(0,x_0)/p_+^*(L,x_0)], 
 \end{equation}
 with $\calJ^*(x_0)=\kappa_0[p_-^*(0,x_0)-p_+^*(L,x_0)]$, see Fig. \ref{fig6}.
 We also find that ${\mathcal I}_{\rm int}^*$ is a nonmonotonic function of the permeability $\kappa_0$, as illustrated in Fig. \ref{fig7}. This follows from the fact that ${\mathcal I}_{\rm int}^*$ vanishes in the limits $\kappa_0\rightarrow 0$ and $\kappa_0\rightarrow \infty$. In the first limit, the circle becomes topologically equivalent to a finite interval with totally reflecting boundaries at both end, whereas in the second limit the interface becomes totally transparent.
 
 \begin{figure}[t!]
\centering
 \includegraphics[width=8cm]{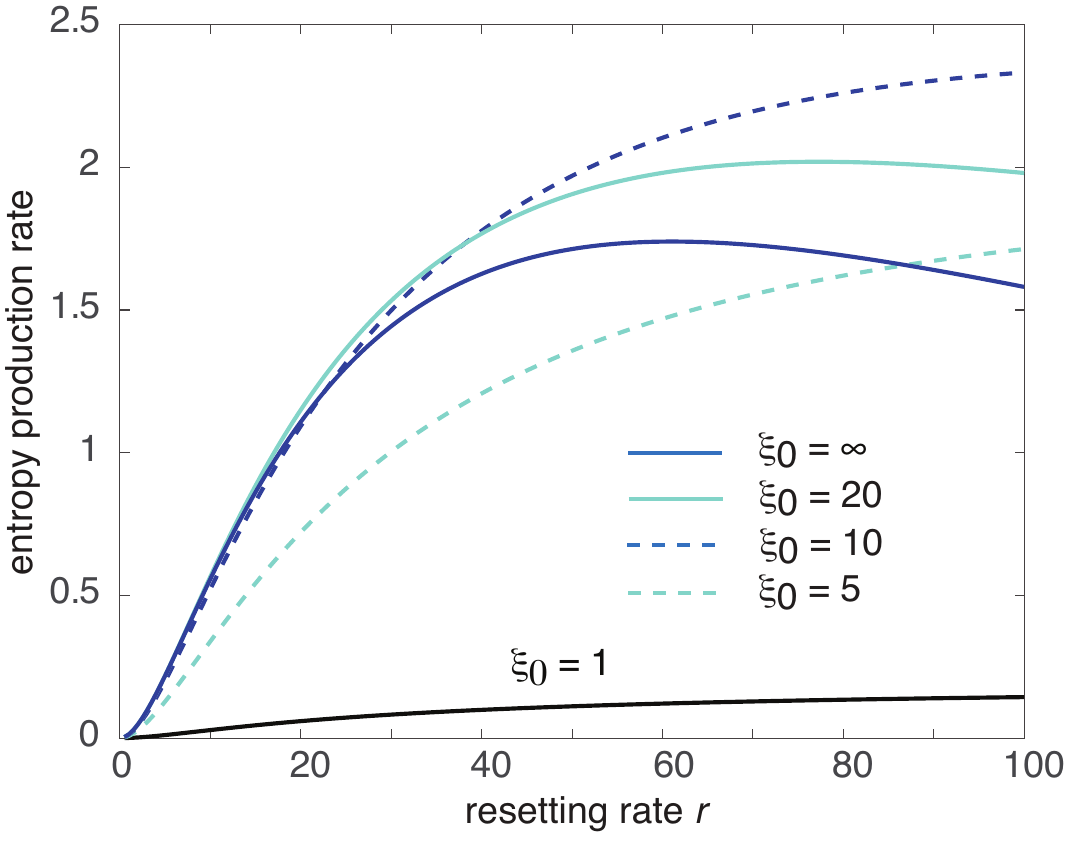}
  \caption{Single-particle diffusion on a ring with resetting to a random location that is distributed according to the probability density (\ref{sig}) parameterized by $\xi_0$. Plots of the average rate of interfacial entropy production ${\mathcal I}_{\rm int}^*$ as a function of $r$ for different values of $\xi_0$. Other parameters are $D=L=1$, $\kappa_0=1$ and $x_0=0.25$.}
  \label{fig8}
\end{figure}

 \begin{figure}[t!]
\centering
 \includegraphics[width=8cm]{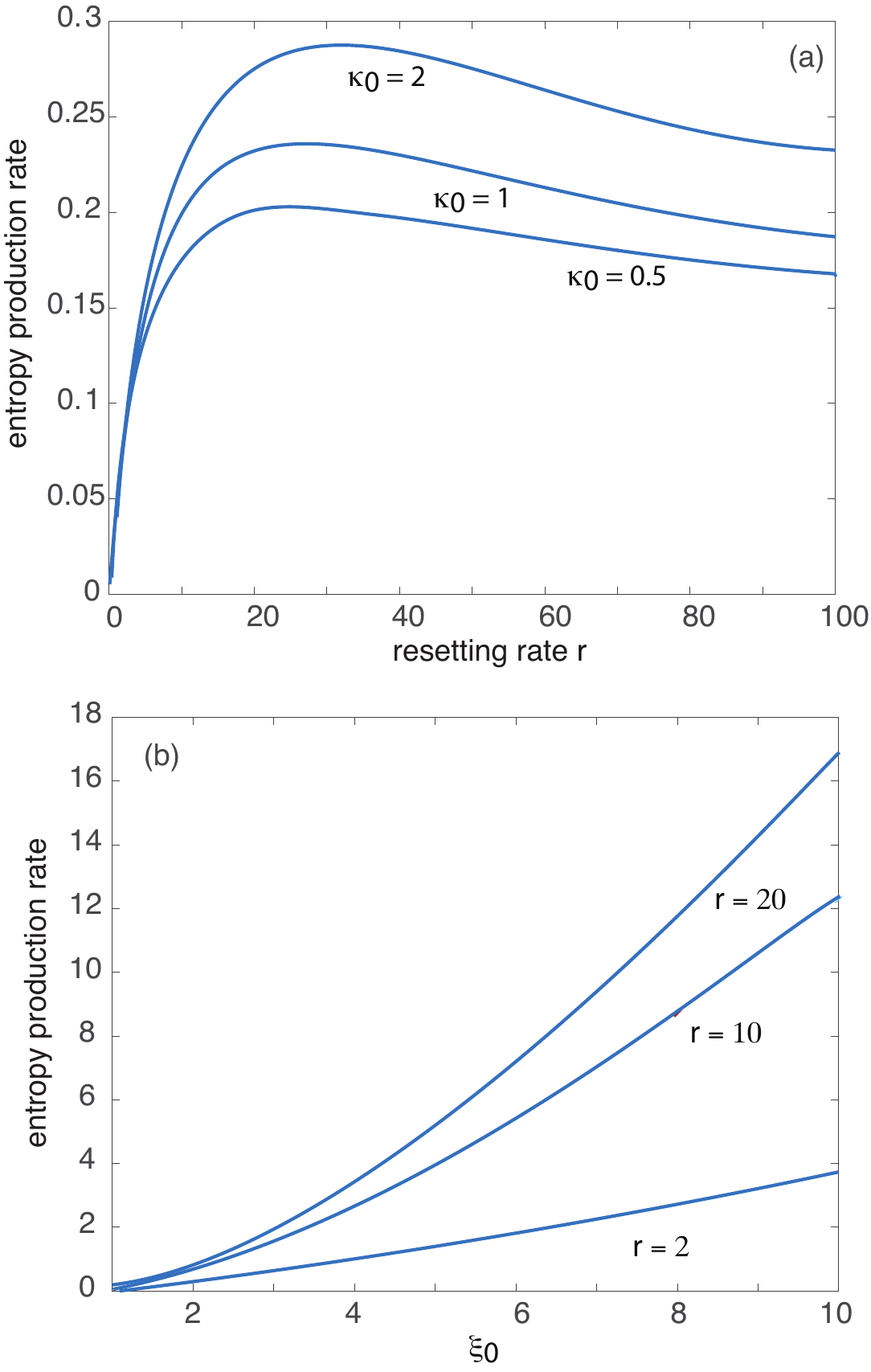}
  \caption{Single-particle diffusion on a ring with resetting to a random location that is distributed according to the probability density (\ref{sig}) parameterized by $\xi_0$. Plots of the average rate of resetting entropy production ${\mathcal I}_{r}^*$ as a function of (a) $r$ for different values of $\xi_0$ and (b) $\xi_0$ for different values of $r$. Other parameters are $D=L=1$, $\kappa_0=1$ and $x_0=0.25$.}
  \label{fig9}
\end{figure}

 Let us now turn to a resetting protocol in which the particle resets to a random point according to the density 
 \begin{subequations}
  \label{sig}
 \begin{align}
 \sigma_0(y,x_0)&=\Lambda_0(x_0) \e^{-|y-x_0|\xi_0}, \\ \Lambda_0(x_0)&=\frac{\xi_0 }{2-\e^{-x_0\xi_0}-\e^{-(L-x_0)\xi_0}}.
 \end{align}
 \end{subequations}
 Note that $\sigma_0(y,x_0)\rightarrow \delta(x-x_0)$ in the limit $\xi_0\rightarrow \infty$ and $\sigma_0(y,x_0)\rightarrow 1/L$ in the limit $\xi_0\rightarrow 0$.
  The corresponding NESS is
 \begin{align}
& p^*(x)\nonumber \\
 &= \int_0^x dy\, p_+(x,y)\sigma_0(y,x_0) +\int_x^L dy\, p_-(x,y)\sigma_0(y,x_0) \nonumber \\
 &={\mathcal A}(x,x_0)\e^{\eta x}+{\mathcal B}(x,x_0)\e^{-\eta x},
 \end{align}
 where
 \begin{subequations}
 \begin{align}
& {\mathcal A}(x,x_0)\\
&=  \int_0^L dy\, A_1(y)\sigma_0(y,x_0)-\int_0^x dy\, \Gamma_-(y)\sigma_0(y,x_0) \nonumber \\
& {\mathcal B}(x,x_0)\\
&= \int_0^L dy\, B_1(y)\sigma_0(y,x_0)+\int_0^x dy\, \Gamma_+(y)\sigma_0(y,x_0).\nonumber
 \end{align}
 \end{subequations}
Moreover,
 \begin{align}
 &\int_0^x dy\, \Gamma_{\pm}(y)\sigma_0(y,x_0)\\
 &\quad =f_{\pm}(x):=\frac{\sqrt{r/D}\Lambda_0\e^{-x_0\xi_0}}{2}\frac{ \left (\e^{x[\pm \eta+\xi_0]}-1\right )}{ \pm \eta+\xi_0}\nonumber
 \end{align}
 for $x<x_0$ and
 \begin{align}
 &\int_0^x dy\, \Gamma_{\pm}(y)\sigma_0(y,x_0)\\
 &\quad = f_{\pm}(x_0)+\frac{\sqrt{r/D}\Lambda_0\e^{x_0\xi_0}}{2}\frac{\left (\e^{x[\pm \eta-\xi_0]}-\e^{x_0[\pm \eta-\xi_0]}\right )}{ \pm \eta-\xi_0}\nonumber
 \end{align}
 for $x>x_0$. In Fig. \ref{fig8} we plot the corresponding interfacial entropy production rate ${\mathcal I}^*_{\rm int}$ as a function of $r$ for different values of the decay parameter $\xi_0$ in the definition of the resetting position density given by Eq. (\ref{sig}). Since $\sigma_0(x)$ becomes a uniform distribution in the limit $\xi_0\rightarrow 0$, it follows that ${\mathcal I}^*_{\rm int}\rightarrow 0$ by symmetry. On the other hand, in the limit $\xi_0\rightarrow \infty$, we recover the curve obtained for resetting to $x_0$. Finally, Eq. (\ref{kk}) implies that for $0<\xi_0<\infty$, there is also a positive contribution to the average entropy production rate from resetting. The stationary version takes the form
\begin{align}
&{\mathcal I}^*_r \\
&=r \int_0^L dx \,[ p^*(x)-\sigma_0(x,x_0)]\ln(p^*(x))-\ln \sigma_0(x,x_0)].\nonumber \end{align}
Example plots of ${\mathcal I}^*_r$ as a function of $r$ and $\xi_0$ are shown in Fig. \ref{fig9}. As expected,  ${\mathcal I}^*_r$ blows up in the limit $\xi_0\rightarrow \infty$ since the corresponding Kullback-Leibler divergences become singular.
\newpage

\setcounter{equation}{0}
\section{Probabilistic interpretation of the entropy contribution ${\mathcal I}_{\rm int}$ }

In Sect. II we analyzed Brownian motion across a semipermeable interface using a Fokker-Planck description of the distribution of sample paths. In order to understand the origins of the interfacial entropy production term ${\mathcal I}_{\rm int}$, see Eq. (\ref{Iint}) and its higher-dimensional analog (\ref{dIint}), we turn to a probabilistic description of individual trajectories based on so-called snapping out Brownian motion \cite{Lejay16,Lejay18,Bressloff22a,Bressloff23}. Here we only cover the essential elements necessary for interpreting ${\mathcal I}_{\rm int}$ in 1D. Extensions of the probabilistic framework to higher-dimensional interfaces can be found in Ref. \cite{Bressloff23}.

    \begin{figure*}[t!]
    \centering
  \includegraphics[width=15cm]{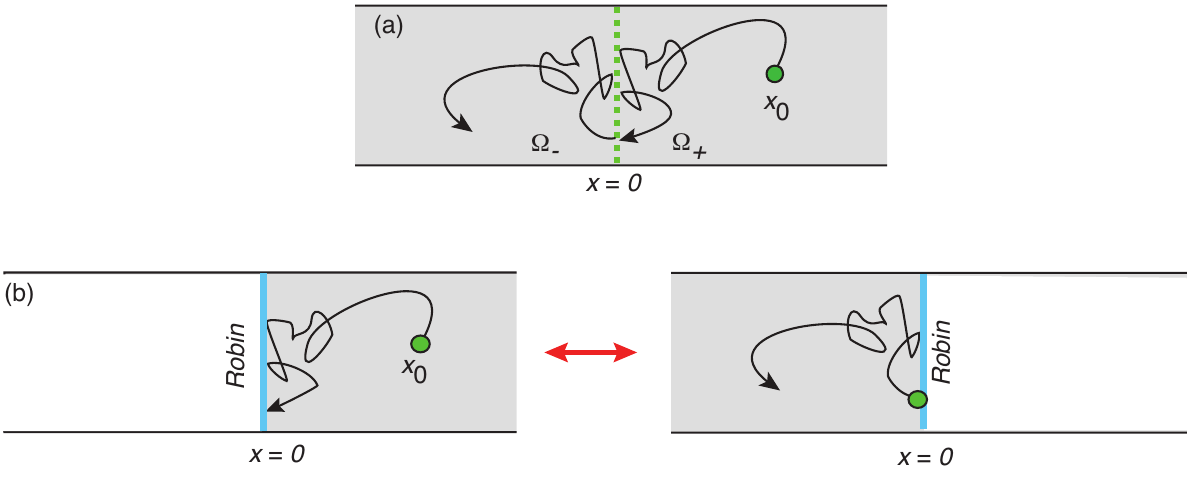}
  \caption{Snapping out BM. (a) Single-particle diffusing across a semipermeable interface at $x=0$. (b) Decomposition of snapping out BM into the random switching between two partially reflected BMs in the domains $\Omega_{\pm}$.}
  \label{fig10}
\end{figure*}

\subsection{Snapping out Brownian motion}

The dynamics of snapping out BM is formulated in terms of a sequence of killed reflected Brownian motions 
    in either $\Omega_-=(-\infty,0^-]$ or $\Omega_+=[0^+,\infty)$ \cite{Lejay16,Lejay18,Bressloff22a,Bressloff23}. Let $\bbT_n$ denote the time of the $n^{\rm th}$ killing (with $\bbT_0 = 0$). Immediately after the killing event, the position of the particle is taken to be
    \begin{align}
        \label{kil}
        X(\bbT_n^+) = \lim_{\epsilon \rightarrow 0^+} \left [-Y_n \epsilon+  (1 - Y_n)\epsilon\right ],
    \end{align}
    where $Y_n$ is an independent Bernoulli random variable with $\P[Y_n=1]=\P[Y_n=0]=1/2$. Suppose that $X(t)\in\Omega_+$ for $t \in (\bbT_{n}, \bbT_{n+1})$, that is, $ X({\bbT_n^+})=0^+$, and introduce the  boundary local time \cite{Levy40,Ito63,Dynkin65,McKean75,Grebenkov06}
    \begin{align}
        \label{local-time-main}
        L_n^+(t)=\lim_{\epsilon\rightarrow 0^+}\frac{D}{\epsilon}\int_0^tI\{0 \leq X(\bbT_n+s)\leq \epsilon\} ds,\quad 
\end{align}
where $I$ is the indicator function. The boundary local time $L_n^{+}(t)$ is a Brownian functional that tracks the amount of the time the particle is in contact with the right-hand side of the interface over the time interval $[\bbT_n,t]$. It can be proven that the local time exists, and is a continuous, positive increasing function of time. The SDE for $X(t)$, $t \in (\bbT_{n}, \bbT_{n+1})$, is given by the so-called Skorokhod equation for reflected BM in the half-line $\Omega_+$: 
    \begin{align}
        \label{sde-outside}
        dX(t)=\frac{1}{\gamma} F(X(t))dt +\sqrt{2D}dW(t)+ dL_n(t)
    \end{align}
    for $ t \in (\bbT_{n}, \bbT_{n+1})$, where $W(t)$ is a Wiener process with $W(0)=0$.
    Formally speaking, 
    \begin{equation}
    dL^+_n(t)= \lim_{\epsilon \rightarrow 0^+} \delta(X(t)-\epsilon)dt,    \end{equation}
so that each time the particle hits the interface it is given a positive impulsive kick back into the domain.
   The time of the next killing is then determined by the condition
    \begin{align}
        \bbT_{n+1} = \bbT_n+\inf\left\{t >0,  \  L_n^+(t)\geq \widehat{\ell} \right\}, 
    \end{align}
where $\widehat{\ell}$ is an independent randomly generated local time threshold with 
    \begin{align}
    \label{Psi}
      \P[\widehat{\ell}>\ell] = \e^{-\kappa_0\ell}, \ \ell \geq 0.
    \end{align}
On the other hand, if $X({\bbT_n^+} )=0^-$ then the next round of reflected BM takes place in the domain $\Omega_-$. The corresponding SDE is
    \begin{align}
        \label{sde-inside}
        dX(t)= \frac{1}{\gamma}F(X(t))dt+\sqrt{2D}dW(t) - dL_n^-(t)
    \end{align}
  with $ t\in (\bbT_n,\bbT_{n+1})$, $ X(t)\in \Omega_-$, 
    \begin{align}
        \label{local-time-j}
      L_n^-(t)=\lim_{\epsilon\rightarrow 0^+}\frac{D}{\epsilon}\int_0^tI\{-\epsilon \leq X(\bbT_n+s) \leq 0\} ds
    \end{align}
and
    \begin{align}
        \bbT_{n+1} =\bbT_n+ \inf\left\{t > \bbT_{n}: \quad  L_n^-(t) \geq \widehat{\ell} \right\}.
    \end{align} 
       
    In summary, the dynamics of snapping out BM consists of sewing together successive rounds of reflected BM, each of which evolves according to the SDE (\ref{sde-outside}) or (\ref{sde-inside}), see Fig. \ref{fig10}. Each round is killed when the local time at the right-hand or left-hand side of the interface exceeds an exponentially distributed random threshold. (The threshold is independently generated each round.) Following each round of killing, an unbiased coin is thrown to determine which side of the interface the next round occurs. It is this randomization that accounts for the term ${\mathcal I}_{\rm int}$ appearing in Eq. (\ref{Rtot}). That is, ${\mathcal I}_{\rm int}$ is given by the product of the flux $\calJ(t)$ across the interface, which specifies the effective rate of randomization, and the corresponding entropy difference $\ln p(0^-,t) -\ln p(0^+,t)$. Of course, this assumes that snapping out BM generates sample paths whose distribution is given by the solution of the corresponding FP Eq. (\ref{rho1D}). We sketch the basic proof based on renewal theory. For an alternative proof in 1D see Ref. \cite{Lejay16} and for the generalization to higher spatial dimensions see Ref. \cite{Bressloff23}.
Let $p(z,t)$ denote the probability density of snapping out BM for $p(x,0)=\delta(x-x_0)$ and $x_0>0$. Let $q(z,t|x_0)$ be the corresponding solution for partially reflected BM in $\Omega_{+}$. (It is straightforward to generalize the analysis to the case of a general distribution of initial conditions $g(x_0)$ that spans both sides of the interface.) The densities $p$ are related to $q$ according to the last renewal equation \cite{Bressloff22a,Bressloff23}
\begin{subequations}
\label{renewal}
 \begin{align}  
& p(x,t)=q(x,t|x_0) \\
 &+\frac{\kappa_0}{2}\int_0^t q(x,\tau|0)[p(0^+,t-\tau ) +p(0^-,t-\tau )]d\tau,\quad x>0,\nonumber  \\
 & p(x,t)\\
 &=\frac{\kappa_0}{2}\int_0^t q(|x|,\tau|0)[p(0^+,t-\tau ) +p(0^-,t-\tau )]d\tau \quad x<0.\nonumber 
   \end{align}
   \end{subequations}
 The first term on the right-hand side of Eq. (\ref{renewal}a) represents all sample trajectories that have never been absorbed by the barrier at $x=0^{\pm}$ up to time $t$. The corresponding integrand represents all trajectories that were last absorbed (stopped) at time $t-\tau$ in either the positively or negatively reflected BM state and then switched to the appropriate sign to reach $x$ with probability 1/2. Since the particle is not absorbed over the interval $(t-\tau,t]$, the probability of reaching $x \in \Omega_+$ starting at $x=0^{\pm}$ is $q(x,\tau|0)$. The probability that the last stopping event occurred in the interval $(t-\tau,t-\tau+d\tau)$ irrespective of previous events is $\kappa_0 d\tau$. A similar argument holds for Eq. (\ref{renewal}b).

The renewal Eqs. (\ref{renewal}) can be used to express $p$ in terms of $q$ using Laplace transforms. First,
\begin{subequations}
  \label{renewal2}
 \begin{align}
 \widetilde{p}(x,s) &= \q(x,s|x_0)+\frac{\kappa_0}{2} \q(x,s|0)[\widetilde{p}(0^+,s )+\widetilde{p}(0^-,s ) ],\nonumber\\
 &\qquad x >0,\\
  \widetilde{p}(x,s) &=\frac{\kappa_0}{2} \q(|x|,s|0)[\widetilde{p}(0^+,s )+\widetilde{p}(0^-,s ) ],
 \quad x <0. 
 \end{align}
 \end{subequations}
  Setting $x=0^{\pm}$ in equation (\ref{renewal2}), summing the results and rearranging shows that
  \begin{equation}
  \label{oo}
 \widetilde{p}(0^+,s)+ \widetilde{p}(0^-,s) =\frac{ \q(0,s|x_0)}{1-\kappa_0\q(0,s|0)} .
 \end{equation}
  Substituting back into equations (\ref{renewal2}) yields the explicit solution
  \begin{subequations}
    \label{renewal3}
 \begin{align}
 \widetilde{p}(x,s) &= \q(x,s|x_0)+ \frac{ \kappa_0\q(0,s|x_0)/2}{1-\kappa_0\q(0,s|0)} \q(x,s|0) ,\quad x >0,\\
 \widetilde{p}(x,s) &=  \frac{ \kappa_0\q(0,s|x_0)/2}{1-\kappa_0\q(0,s|0)} \q(|x|,s|0),\quad x <0.
 \end{align}
 \end{subequations}
 Calculating the full solution $p(x,t)$ thus reduces to the problem of finding the corresponding  solution $q(x,t|x_0)$ of partially reflected BM in $\Omega_{+}$. As we have shown elsewhere, this then establishes that $p(x,t)$ satisfies the interfacial conditions (\ref{rho1D}c). 
 
Finally, note that it is also possible to incorporate the directional asymmetry of Eq. (\ref{asym}) into snapping out BM \cite{Bressloff23}. This is achieved by introducing a bias in the switching between the positive and negative directions of reflected BM following each round of killing. More specifically, the independent Bernoulli random variable $Y_n$ in Eq. (\ref{kil}) now has the biased probability distribution $\P[Y_n=0]=\alpha$ and $\P[Y_n=1]=1-\alpha$ for $0<\alpha <1$.
The 1D renewal Eq. (\ref{renewal2}) becomes
\begin{subequations}
  \label{bias2}
 \begin{align}
 \widetilde{p}(x,s) &= \q(x,s|x_0)+\frac{\kappa_0 \alpha }{2} \q(x,s|0)[\widetilde{p}(0^+,s )+\widetilde{p}(0^-,s ) ],\nonumber \\
 &\qquad x>0\\
  \widetilde{p}(x,s) &=\frac{\kappa_0[1-\alpha]}{2} \q(|x|,s|0)[\widetilde{p}(0^+,s )+\widetilde{p}(0^-,s ) ] ,\quad x <0.
 \end{align}
 \end{subequations}
 Setting $x=0^{\pm}$ in Eqs. (\ref{bias2}), summing the results and rearranging recovers Eq. (\ref{oo}).
It can then be shown that  snapping out BM with biased switching and $\alpha >1/2$ is equivalent to single-particle diffusion through a directed semipermeable barrier with an effective permeability $\kappa_0 \alpha/2$ and bias $\sigma =(1-\alpha)/\alpha$.

\subsection{Stochastic entropy production}

The probabilistic formulation of snapping out BM can also be used to construct a stochastic version of entropy production. For the sake of illustration, we focus on the 1D unbiased case. Let $X(t)$ be the position of the Brownian particle at time $t$ and consider the stochastic system entropy 
\begin{equation}
\label{sent}
S^{\rm sys}(t)=-\ln p(X(t),t),
\end{equation}
where $p(x,t)$ is the solution of Eqs. (\ref{rho1D}).
Differentiating both sides of Eq. (\ref{sent}) with respect to $t$ and using the chain rule in the Stratonovich version of stochastic calculus gives
\begin{widetext}
\begin{align}
\label{dS}
 dS^{\rm sys}(t)&=-\frac{1}{ p(X(t),t)}\frac{\partial  p(X(t),t)}{\partial t}dt  -\frac{1}{ p(X(t),t)}\frac{\partial  p(X(t),t)}{\partial x}\circ dX(t) \\
&\quad -\frac{\kappa_0}{2}\lim_{\epsilon \rightarrow 0^+}[\delta(X(t)-\epsilon)-\delta(X(t)+\epsilon)]\ln [p(0^-,t)/ p(0^+,t)]dt ,\nonumber 
\end{align}
with
\begin{align}
dX(t)&=-\frac{1}{\gamma} V'(X(t))dt +\sqrt{2D}dW+D[1-\kappa_0 dt/2]\lim_{\epsilon \rightarrow 0^+}[\delta(X(t)-\epsilon)-\delta(X(t)+\epsilon)]dt.
\label{DelX}
\end{align}
\end{widetext}
The second-term on the right-hand side of Eq. (\ref{dS}) takes into account the entropy change due to an infinitesimal change in particle position that does not cross the interface. Hence, the Dirac delta functions in Eq. (\ref{DelX}) are the local time impulses that reflect the particle at $x=0^{\pm}$, and the factors $[1-\kappa_0 dt/2]$ represent the probability that reflection occurs.
The third term on the right-hand side of Eq. (\ref{dS}) can be understood as follows. If the particle is located at $x=0^{\pm}$ then the probability of the reflected BM being killed in the interval $[t,t+dt]$ is $\kappa_0 dt$, after which there is a probability of 1/2 that it switches to the other side of the interface. This results in an entropy change equal to $\Delta S^{\rm sys}=\pm \ln [p(0^-,t)/ p(0^+,t)]$, whose sign depends on the crossing direction. A similar interpretation applies to the fourth term. In terms of the probability flux $J(x,t)$, we have
\begin{widetext}
\begin{align}
& -\frac{1}{ p(X(t),t)}\frac{\partial  p(X(t),t)}{\partial x}\circ dX(t) =\frac{J(X(t),t)}{ Dp(X(t),t)}\circ dX(t)-dS^{\rm env}(X(t),t),\quad dS^{\rm env}(X(t),t)=-\frac{V'(X(t))}{T}\circ dX(t),
\label{stot}
\end{align}
 where $dS^{\rm env}(X(t),t) $ is the the infinitesimal change in the environmental entropy.
In order to determine the average entropy production rate we need to take expectations with respect to the white noise process. This is simplified by converting from Stratonovich to It\^o calculus \cite{Peliti21}. In particular, to leading order in $dt$,
\begin{align*}
  \frac{J(X(t),t)}{D p(X(t),t)}\circ dX  
&=-  \frac{J(X(t),t)V'(X(t))}{T p(X(t),t)}  dt+\sqrt{\frac{2}{D}}\frac{J(X(t),t)}{ p(X(t),t)}\circ dW  + \frac{J(X(t),t)}{Dp(X(t),t)}\lim_{\epsilon \rightarrow 0^+}[\delta(X(t)-\epsilon)-\delta(X(t)+\epsilon)]dt\nonumber \\
&=-  \frac{J(X(t),t)V'(X(t))}{T p(X(t),t)}  dt+  \sqrt{\frac{2}{D}}\frac{J(X(t),t)}{ p(X(t),t)}\cdot dW   + \frac{1}{ p(X(t),t)} \frac{\partial J(X(t),t)}{\partial x} \ dt\\
&\quad - \frac{J(X(t),t)}{ p^2(X(t),t)}\frac{\partial p(X(t),t)}{\partial x}   dt ,
\end{align*}
since $J(0^-,t)=J(0^+,t)$. Hence, we can rewrite Eq. (\ref{stot}) as
\begin{align}
\label{stot2}
 dS^{\rm sys}(X(t),t) +dS^{\rm env}(X(t),t) 
 &=-\frac{1}{ p(X(t),t)}\left [\frac{\partial  p(X(t),t)}{\partial t}- \frac{\partial J(X(t),t)}{\partial x}\right ]dt+  \sqrt{\frac{2}{D}}\frac{J(X(t),t)}{ p(X(t),t)}dW\\
&\quad + \frac{J^2(X(t),t)}{D p^2(X(t),t)} dt -\frac{\kappa_0}{2}\lim_{\epsilon \rightarrow 0^+}[\delta(X(t)-\epsilon)-\delta(X(t)+\epsilon)]\ln [p(0^-,t)/ p(0^+,t)]dt.\nonumber
\end{align}
\end{widetext}
We now average each term in Eq. (\ref{stot2}) with respect to the white noise  process using the following identity for any integrable function $g(X(t))$:
\begin{align}
 \left \langle g(X(t)\right \rangle &=  \left \langle \fint dx\, \delta(x-X(t) )g(x)\right \rangle  \\
&=\fint dx \, g(x)   \left \langle  \delta(x-X(t) \right \rangle=\fint dx \, g(x)p(x,t).\nonumber 
\end{align}
First,
\begin{align}
&\left \langle\frac{1}{ p(X(t),t)}\frac{\partial  p(X(t),t)}{\partial t} \right \rangle\nonumber \\
& \quad =\fint dx \frac{\partial  p(x,t)}{\partial t} =\frac{\partial  }{\partial t}\fint dx \,p(x,t)=0
\end{align}
by conservation of probability. Second,
\begin{align}
&\left \langle\frac{1}{ p(X(t),t)}\frac{\partial  J(X(t),t)}{\partial x} \right \rangle\nonumber 
\\ &  \quad  =\fint dx \frac{\partial  J(x,t)}{\partial x} =J(0^-,t)-J(0^+,t)=0.
\end{align}
Taking expectation of the Wiener process in the It\^o product terms also gives zero. 
Hence, combining the various results recovers Eq. (\ref{Rtot}), indicating that we can reverse the order of integration and differentiation so that
\begin{align}
\left \langle \frac{dS^{\rm sys}(X(t),t)}{dt}\right \rangle=\frac{d}{dt}\bigg \langle S^{\rm sys}(X(t),t)\bigg \rangle \end{align}
etc.

\setcounter{equation}{0}
\section{Discussion}

In this paper we showed how the presence of a semipermeable interface $\calS$ increases the average rate of entropy production of a single diffusing particle by an amount that is equal to the product of the flux through the interface and the logarithm of the ratio of the probability density on either side of the interface, integrated along $\calS$. We illustrated the theory using the example of diffusion with stochastic resetting on a circle, and showed that the average rate of interfacial entropy production in the NESS is a nonmonotonic function of the resetting rate and the permeability. Finally, we presented a probabilistic interpretation of the interfacial entropy production rate that is based on snapping out BM. The latter represents individual stochastic trajectories as sequences of partially reflected BMs that are restricted to one side of the interface or the other. When a given round of partially reflected BM is terminated, a Bernoulli random variable is used to determine which side of the interface the next round takes place. We identified this switching process as the source of interfacial entropy production. Moreover, we showed how a biased switching process is equivalent to a directionally biased interface arising from a jump discontinuity in the chemical potential. The latter contributes to the dissipation of heat into the environment.
Snapping out BM also allowed us to construct a stochastic version of entropy production defined along individual trajectories. Averaging with respect to the distribution of trajectories recovered the expression for the average rate of entropy production obtained from the Gibbs-Shannon entropy.

One direction for future work would be to relate the stochastic entropy for diffusion through a semipermeable interface to the ratio of forward and backward path probabilities \cite{Seifert12,Peliti21}. More specifically, a fundamental result of stochastic thermodynamics is that for many continuous stochastic processes, the total stochastic or instantaneous entropy production can be expressed as
\begin{equation}
S^{\rm tot}(t)= \ln \left [\frac{\calP[X(\tau),0\leq \tau\leq t]}{\calP[X(t-\tau),0\leq \tau\leq t] }\right ].
\label{path1}
\end{equation}
Such a relationship provides a basis for deriving a variety of fluctuation relations \cite{Seifert12,Peliti21}. 
The corresponding average production rate in steady state is then
\begin{equation}
\calR^{\rm tot}=\lim_{t \rightarrow \infty}\frac{1}{t}\left \langle  \ln \left [\frac{\calP[X(\tau),0\leq \tau\leq t]}{\calP[X(t-\tau),0\leq \tau\leq t] }\right ]\right \rangle.
\label{path2}
\end{equation}
One way to establish a result of the form (\ref{path1}) is to use path integrals. Within a path integral framework, one could treat diffusion as a random walk on a lattice in which a semipermeable barrier is represented in terms of  local defects \cite{Powles92,Kenkre08,Novikov11,Kay22}. 

Finally, note that in this paper we considered a mesoscopic model of diffusion through a semipermeable interface, which involved phenomenological parameters such as the permeability $\kappa_0$ and the directional bias $\sigma$. These also appeared as parameters in snapping out BM, with $\kappa_0$ determining the rate at which each round of partially reflected BM is killed and $\sigma$ specifying the bias of the switching Bernoulli process. Another direction of future work would be to develop a microscopic model of a semipermeable interface that identifies the biophysical mechanisms underlying $\kappa_0$, $\sigma$, and interfacial entropy production. Along these lines, we have recently proposed a more general model of snapping out BM, in which each round of partially reflected BM is killed according to a more general threshold distribution than the exponential (\ref{Psi}) \cite{Bressloff22a,Bressloff23}. The corresponding effective permeability becomes a time-dependent function that can have heavy tails.


\end{document}